\documentclass[a4paper]{panl}
\usepackage{cite}
\usepackage{wrapfig}
\usepackage{graphicx}
\usepackage{amssymb}
\usepackage{amsfonts}
\usepackage{amsmath}
\usepackage{longtable}
\usepackage{rotating}
\usepackage{lscape}
\usepackage{epsfig}
\usepackage{multirow}
\usepackage[normalem]{ulem} 

\originalTeX
\begin{document}
% Journal sections (see http://pkp.jinr.ru/index.php/PEPAN_LETTERS/about/editorialPolicies#focusAndScope)
\issuearea{Physics of Elementary Particles and Atomic Nuclei. Theory}
% or in Russian
%\issuearea{ФИЗИКА ЭЛЕМЕНТАРНЫХ ЧАСТИЦ И АТОМНОГО ЯДРА. ТЕОРИЯ}

\title{Mott dissociation and kaon to pion ratio in the EPNJL model}
%\title{The medium effects on the kaon to pion ratio in the frame of effective QCD models}
\maketitle
\authors{D. Blaschke$^{a,b,c}$\footnote{E-mail: david.blaschke@gmail.com}, A. V. Friesen$^{a,d}$\footnote{E-mail: avfriesen@theor.jinr.ru}, Yu. L. Kalinovsky$^{a,d}$\footnote{E-mail: kalinov@jinr.ru}, A. E. Radzhabov$^{e}$\footnote{E-mail: andrey.radjabov@yandex.ru}}
\from{$^{a}$\,Joint Institute for Nuclear Research, 141980, Dubna, Russia}
\vspace{-3mm}
\from{$^{b}$\,Institute for Theoretical Physics, University of Wroc{\l}aw, 50-204 Wroc{\l}aw, Poland}
\vspace{-3mm}
\from{$^{c}$\,National Research Nuclear University (MEPhI), 115409 Moscow, Russia}
\vspace{-3mm}
\from{$^{d}$\,Dubna State University, 141982, Dubna, Russia}
\vspace{-3mm}
\from{$^{e}$\,Matrosov Institute for System Dynamics and Control Theory, Irkutsk 664033, Russia}

\begin{abstract}
% Russian translation of the abstract
The behaviour of pseudoscalar mesons within the SU(3)PNJL-like models is considered for finite T and $\mu_B$.  We compare the pole approximation (Breit-Wigner) with the Beth-Uhlenbeck approach. We evaluate the $K/\pi$ ratios along the phase transition line in the T-$\mu_B$ plane with constant and $T/\mu_B$-dependent pion and strange quark chemical potentials.  Using the model, we can show that the splitting of kaon and anti-kaon masses appears as a result of introduction of density and this explains the difference in the $K^+/\pi^+$ ratio and $K^-/\pi^-$ ratio at low $\sqrt{s_{NN}}$ and their tendency to the same value at high $\sqrt{s_{NN}}$. A sharp "horn" effect  in the $K^+/\pi^+$ ratio is explained by the enhanced pion production which can be described by occurrence of a nonequilibrium pion chemical potential of the order of the pion mass. We elucidate that the horn effect is not related to the existence of a critical endpoint in the 
QCD phase diagram.
\end{abstract}
\vspace*{6pt}


\label{sec:intro}
\section*{Introduction}

Our interest to the ''horn'' in the $K^+/\pi^+$ ratio (the peak-like structure at the collision energy $\sqrt{s_{NN}} \sim 8$ GeV) appears due to the fact  that modern attempts to explain its existence give good intuitive explanation of the $K/\pi$ bechaviour in high-energy region. The smooth, almost constant behaviour  at high energies (RHIC, LHC) can be explained by the fact that at high energies during the heavy ion collision the quark-gluon-plasma is created (it is confirmed by other signals). The creation of the QGP is accompanied by the constant, independent on the temperature yield of the strange particles. The jump from the maximum value to the almost constant line is connected with the increase of the pion yield in comparison with strangeness yield when deconfinement appears and it could be a signal of ''onset of deconfinement'' (the idea was originally predicted by Gazdzicki and Gorenstein) \cite{Gazdzicki:2020jte}.

The description of the $K^+/\pi^+$ at energies $< 8$  GeV has not such a clear explanation. From one hand it was shown that the partial chiral symmetry restoration could be responsible for the quick increase in the $K^+/\pi^+$ ratio at low energies and its decrease with increasing energy (as a result of chiral condensate destruction) \cite{Palmese:2016rtq,Moreau:2017dgq}. From other hand it is not clear still  if the deconfinement appeared during heavy ion collision  at such energies and if it coincided with the chiral phase transition. And at last, the recent data from the NA61 experiment have shown a strong dependence on the system size with no horn effect in Ar+Sc collisions \cite{Gazdzicki:2020jte} for which the reason is not yet understood.

The main idea of this work is find the model able to reproduce both chiral phase transition and deconfinement transition at finite T and $\mu_B$. From this point the NIL-like models are promising models, as the models are capable of describing both the chiral phase transition and deconfinement transition. The chiral symmetry breaking is referred to the quarks developing quasiparticle masses by propagating in the chiral condensate and the confinement of coloured quark states is effectively taken into account by coupling the chiral quark dynamics to the Polyakov loop and its effective potential. However, in the model the absolute value of the pseudocritical temperature of the chiral crossover transition with above 200 MeV is too large when compared to the lattice QCD result for 2+1 flavors of $T_c=156.5\pm 1.5$ MeV \cite{Bazavov:2018mes}.

The so-called entanglement PNJL (EPNJL) model, with the modified scalar four-quark interaction $g_S$ could be one of possible solutions of this problem \cite{Sakai:2010rp,Ruivo:2012xt}.  Such entanglement leads to a change of the phase diagram: the (pseudo-)critical temperature of the chiral and deconfinement crossover transition at low chemical potentials is lowered towards the value of the Lattice QCD prediction. More physical solution is to go beyond the mean field approximation and to consider the role of hadronic excitations in the medium in melting the chiral condensate  \cite{Jankowski:2012ms}. First steps in this direction have been explored, e.g., within a generalized Beth-Uhlenbeck approach \cite{Blaschke:2016fdh,Blaschke:2017pvh}. We also considered the (E)PNJL model with the vector interaction to evaluate the effect of changing the chiral phase transition of the first order to the soft crossover in the region of high $\mu_B$. The including of the vector interaction to the model gives such possibility as the position of the critical end point (CEP) in the phase diagram depends on the value of the coupling $g_V$. When $g_V$ reaches the critical value, the chiral phase transition turns into soft crossover overall \cite{Stephanov:2007fk,Friesen:2014mha}.

In the work we show, that the energy-dependent behaviour of the $K\pi$ ratio could be rescaled by new variable $T/\mu_B$, where both T and $\mu_B$ are chosen along the line of the chiral phase transition, which is supposed to correspond to the chemical freeze-out line in the QCD phase diagram. In this region can be important to introduce the nonequilibrium  pion chemical potential. The  reason to consider the nonequilibrium pion distribution function lies in the behaviour of the pions during the heavy ion collision: their number is quasi conserved over the time scale of the HIC until freeze-out. The resulting nonequilibrium pion distribution function with the pion chemical potential has been successfully used in the phenomenological description of the transverse momentum spectra of pions produced
in heavy-ion collision experiments, see \cite{Begun:2013nga}.

\label{sec:model}
\section*{Mass spectrum for mesons at finite temperature and density}

We start from the  SU(3) PNJL model with the Kobayashi - Maskawa - t'Hooft (KMT) interaction \cite{Fukushima:2008wg}
\begin{eqnarray}
\mathcal{L\,} & = & \bar{q}\,(\,i\,{\gamma}^{\mu}D_{\mu}-\hat{m} - \gamma_0\hat{\mu})\,q 
+  \frac{1}{2}g_{S}\,\sum_{a=0}^{8}\,[\,{(\,\bar{q}\,\lambda^{a}\,q\,)}^{2}+{(\,\bar{q}\,i\gamma_{5}\,\lambda^{a}\,q)}^{2}\,] \nonumber \\
&& + g_D \left\{\mbox{det}\,[\bar{q}\,(1+\gamma_{5})\,q\,]+\mbox{det}\,[\bar{q}\,(1-\gamma_{5})\,q\,]\right\} -  \mathcal{U}(\Phi, \bar{\Phi}; T). 
\label{lagr}%
\end{eqnarray}

The model describes the dynamical chiral symmetry breaking as a coupling of quarks to the chiral condensate. The quarks develop a quasiparticle mass even for vanishing current quark masses $m_{0i}$ (chiral limit)  by propagating in the chiral condensate. In this model with Lagrangian (\ref{lagr})  the quarks are coupled both to the chiral condensate and to the homogeneous gluon background fields representing by the Polyakov loop dynamics. The confinement properties are described by the effective potential $\mathcal{U}(\Phi, \bar{\Phi}; T)$, which depends on the complex traced Polyakov loop $\Phi={\rm Tr}_c \exp[i\beta (\lambda_3 A_3^0+\lambda_8A_8^0)]$ and its conjugate $\bar{\Phi}$. The potential is constructed on the basis of the center symmetry $Z_3$ of the color SU(3) gauge group. The possible temperature dependence of its parameters is fitted to lattice QCD results for the pressure in the pure gauge sector (for details see \cite{Ratti:2005jh}). 

%For the numerical calculations in this work we used the following set of parameters:  the current quark masses $m_{0 u} = m_{0 d} = 5.5$ MeV, $m_{0 s} = 0.131$ GeV,  the cut-off $\Lambda = 0.652$ GeV and the couplings 
%$g_D = 89.9$ GeV$^{-5}$, $g_S = 4.3$ GeV$^{-2}$. 
%$g_D \Lambda^5 = 10.592$, $g_S \Lambda^2 = 1.828$. We also introduced the strange quark chemical potential as $\mu_s = 0.55\mu_u $ and $\mu_u = \mu_d$ \cite{Friesen:2018ojv}. For effective potential $\mathcal{U}(\Phi, \bar{\Phi}; T)$ we used  the standard polynomial form \cite{Ratti:2005jh}, where the original parameter $T_0=270$ MeV for the deconfinement transition temperature in pure gauge was rescaled to $T_0=187$ MeV for the system with $N_f=2+1$ flavors \cite{Schaefer:2007pw}. 

The pseudocritical temperature of the chiral crossover transition at $\mu_B = 0$ GeV in the mean field PNJL model is $T_c = 0.218$ GeV  and this value is higher than the $T_c =156.5\pm 1.5$ GeV  obtained  by the ab-initio simulations of
Lattice QCD \cite{Bazavov:2018mes}. 

PNJL model with rescaled $T_0$ has a problem with correspondence between critical temperature for the chiral condensate and for the Polyakov loop. To improve this disagreement with Lattice data, was obtained so-called EPNJL model  \cite{Ruivo:2012xt}.
According to this extension of the PNJL model, a phenomenological dependence of the scalar meson coupling $g_S$ on the Polyakov loop is introduced that does obey the $Z_3$-symmetry 
\begin{equation}
\tilde{g}_S(\Phi) = g_S(1-\alpha_1\Phi\bar{\Phi} - \alpha_2(\Phi^3 + \bar{\Phi}^3)),
\label{Gt}
\end{equation}
where the parameters $\alpha_1 = \alpha_2 = 0.2$ were chosen in \cite{Sakai:2010rp} to reproduce the two-flavor LQCD data.  Such rescaling of $g_S$ leads to a rescaling of the pseudocritical temperature in the low-density region \cite{Sakai:2010rp,Ruivo:2012xt} to $T_c\sim 0.179$ GeV and now being synchronous with the Polyakov-loop transition.

We also considered the PNJL model with vector interaction to discuss the critical end point position on the phase diagram. The position of  the CEP is under debate \cite{Stephanov:2007fk} and can not be settled yet by Lattice QCD simulations. It was shown in the NJL and in the PNJL modes  with vector interaction, that the CEP can disappear when the vector coupling exceeds a critical value. The Lagrangian of the PNJL model with vector interaction is obtained from the Lagrangian (\ref{lagr}) by adding the vector interaction contribution $\displaystyle{ - \frac{1}{2}g_{V} \sum_{a=0}^{8}\,\,{(\,\bar{q}\gamma_\mu\lambda^{a}\,q\,)}^{2}}$
%\begin{eqnarray}
%\mathcal{L}_{\rm V}  =  - \frac{1}{2}g_{V} \sum_{a=0}^{8}\,\,{(\,\bar{q}\gamma_\mu\lambda^{a}\,q\,)}^{2},
%\label{lagr_v}%
%\end{eqnarray}
the vector coupling $g_V$ can also be rescaled in the spirit of the EPNJL model, similar to the scalar coupling  (\ref{Gt}).
%,
%\begin{equation}
%\tilde{g}_V(\Phi) = g_V(1-\alpha_1\Phi\bar{\Phi} - \alpha_2(\Phi^3 + \bar{\Phi}^3)),
%\label{Gvt}
%\end{equation}
%with the same parameters as in Eq.~(\ref{Gt}).

The meson spectral properties are encoded in the meson propagator (M=P, S)
\begin{equation}
\mathcal{S}_{ij}^M (\omega,\mathbf{q}) = \frac{2 P_{ij}}{1 - 4 P_{ij}\Pi_{ij}^{M}(\omega,\mathbf{q})}~, \label{rprop}
\end{equation}
where for non-diagonal pseudo-scalar mesons $\pi\,,K$:
\begin{eqnarray}
P_{ij}^{\pi}&=&g_{S}+g_{D}\left\langle\bar{q}_{s}q_{s}\right\rangle, \ \ \ 
P_{ij}^{K} = g_{S}+g_{D}\left\langle\bar{q}_{u}q_{u}\right\rangle \label{Pkaon}
\end{eqnarray}
and $\Pi_{ij}^{M}(P_{0})$ is the polarization function. 
%To take into account the complexity of the polarization propagator, 
Particularly interesting are the (complex) mass pole solutions $z_M=m_M - i \Gamma_M/2$ in the Breit-Wigner (BW) approximation of the  Bethe-Salpeter equation
at vanishing meson momentum $\mathbf{q}=0$, 
\begin{equation}
1 - 2 P_{ij}\Pi_{ij}^{M}(z=z_M,\mathbf{q=0}) =0, 
\label{rdisp}
\end{equation}
where for the width of the quark-antiquark resonant state $\Gamma_M \ll m_M$ should hold. 
When the mass parameter is below the two-quark threshold ($m_M < m_i + m_j$), the polarization function is real and Eq. (\ref{rdisp}) corresponds to the homogeneous Bethe-Salpeter equation in the rest frame of the meson which defines a true bound state with mass $m_M$ and infinite lifetime ($\Gamma_M=0$). 
$\Gamma_M$ becomes finite at the Mott transition, when the meson becomes unbound and its mass is larger than that of its constituents 
$m_M > m_i+m_j$. 
The temperatures of the Mott transition for pion and kaon are $T^\pi_{\rm Mott} = 0.232$ GeV and $T^K_{\rm Mott} = 0.230$ GeV, respectively.

An adequate account for both, bound and continuum states can be made within the Beth-Uhlenbeck (BU) approach that is generalized for the case of dense matter 
and the Mott dissociation of mesons across the chiral/deconfinement transition in \cite{Wergieluk:2012gd,Blaschke:2013zaa}. 
For this, the mesonic propagator can be rewritten in the ''polar'' representation:
\begin{equation}
\mathcal{S}^{M}_{ij}(\omega,\bar{q})= |\mathcal{S}^{M}_{ij}(\omega,\bar{q})|e^{i\delta_M(\omega,\bar{q})}
\label{polar}
\end{equation}
with meson phase shift $\delta_M(\omega,\bar{q})$.
%\begin{equation}
%\delta_M(\omega,\bar{q}) = -{\rm{arctan}}\left\{\frac{{\rm{Im}}([{\mathcal{S}}^{M}_{ij}(\omega-i\eta,\bar{q})]^{-1})}{{\rm{Re}}([\mathcal{S}^{M}_{ij}(\omega+i\eta,\bar{q})]^{-1})}\right\}.
%\label{phaseshift}
%\end{equation}
In the phase shift representation, the bound state appears at the energy where the phase shift has a jump to the value $\pi$ \cite{Wergieluk:2012gd}. 
In the rest frame of the meson this energy corresponds to its mass. 
For energies above the continuum threshold, the phase shift drops and asymptotically vanishes,  in accordance with the Levinson theorem.

%The phase shift for the pion and charged kaons for $T= 90$ MeV and nonstrange chemical potential $\mu=350$ MeV are shown in the Fig. \ref{Fig:masses}.
%\begin{figure}[t]
%\begin{center}
%\includegraphics[width=58mm]{Delta} 
%\includegraphics[width=63mm]{masses_n_.png} 
%\vspace{-3mm}
%\caption{Figure caption}
%\end{center}
%\labelf{Fig_masses}
%\vspace{-5mm}
%\end{figure}

At zero chemical potential both the BW and the BU approach show the degeneracy of positive and negative charged meson masses, as for light quarks $m_u=m_d$ was chosen. At nonzero chemical potential and low T, the splitting of mass in charged multiplets appears due to excitation of the Dirac sea modified by the presence of the medium (see \cite{Friesen:2018ojv,Costa:2003uu,Blaschke:2020bzh}). 
In BU approach an anomalous mode appears after $T_{\rm Mott}$ for positively charged kaons. 
This is due to the fact that with increasing density the scattering mode of the polarization function $\Pi_{ij}^{M}(P)$ becomes dominant  and induces a bound state pole in the meson propagator (\ref{rprop}) \cite{Dubinin:2016wvt}.

\label{sec:horn}
\section*{Kaon to pion ratio}

In this Section we discuss the application of the PNJL model to description of the kaon to pion ratio. 
There are some remarks to our calculations in the frame of the PNJL model.

The first point is that experimental data are shown as a function of the collision energy $\sqrt{s_{NN}}$ which never appears as a parameter in effective models. 
We used the fact that in the statistical model for each experimental energy of collision the temperature and the baryon chemical potential of freeze-out can be found using parametrization suggested, e.g., by Cleymans et al. \cite{Cleymans:2005xv}.
%\begin{eqnarray}
%T(\mu_B) &=& a - b\mu^2_B- c\mu^4_B,\label{param_exp} \\
%\mu_B(\sqrt{s})& = &\frac{d}{1 + e\sqrt{s}},
%\label{param_exp1}
%\end{eqnarray}
%where $a = 0.166 \pm 0.002$ GeV, $b = 0.139 \pm 0.016$ GeV$^{-1}$,
%and $c = 0.053 \pm 0.021$ GeV$^{-3}$, $d = 1.308 \pm 0.028$ GeV, $e = 0.273 \pm 0.008$ GeV$^{-1}$.  
Using this parametrization, the $K/ \pi$ ratio can be considered as a function of a new variable $T/\mu_B$ instead of $s_{NN}$, where (T, $\mu_B$) are taken along the freeze-out line.

The second point is that according to the nonequilibrium statistical models the ratio of the yields of mesons, such as the $K/\pi$ ratios which were obtained in the midrapidity range, can be calculated in terms of the ratio of the number densities of mesons ($K^\pm/\pi^\pm = n_{K^\pm}/n_{\pi^{\pm}}$) with
\begin{eqnarray}
\label{eq:nM}
n_{M} &=& d_M \int_0^\infty \frac{\rm d^3 {\bf q}}{(2\pi)^3}g_M(E_M),
\end{eqnarray}
where $g_M(E) = (e^{(E-\mu_M)/T}-1)^{-1}$ is the Bose function for a meson with energy $E$ and chemical potential $\mu_M$.
In the generalized BU approach, however, an off-shell generalization of the partial number density of the mesonic species $M$ holds
\begin{eqnarray}
\label{eq:BU}
n_M(T)&=& 
d_M\int\frac{{ d}^3 \mathbf{q}}{(2\pi)^3} \int \frac{d\omega}{2\pi} g_M(\omega)\frac{d\delta_{M}(\omega,\mathbf{q} )}{d\omega},
%\nonumber 
%\\
%&=& \frac{d_{\rm M}}{T} \int\frac{{d}q\, q^2}{2\pi^2}\int_0^\infty\frac{{d}{\omega}}{2\pi} g_M(\omega )(1+g_M(\omega))\delta_M({\omega}),
\label{eq:BUpartial}
\end{eqnarray}
where  $\delta_M(\omega)=\delta_{M}(\omega,\mathbf{0})$ is the meson phase shift that was calculated for mesons at rest, but 
for which Lorentz-boost invariance $\omega=\sqrt{q^2+{m}^2}$  was assumed, see \cite{Blaschke:2013zaa}.

The chemical potential for pions as a parameter characterizing the nonequilibrium state has been chosen as a constant close to the pion mass, e.g., $\mu_\pi = 0.135$ GeV, following the works \cite{Begun:2013nga}. The chemical potential for kaons is defined here by their chemical composition, i.e. by their quark content, so that $\mu_{K^+} =\mu_u-\mu_s$ and $\mu_{K^-} =\mu_s-\mu_u= - \mu_{K^+}$.

%\begin{figure}[h]
%\begin{center}
%\includegraphics[width = 60mm]{pd_with_freezeout_dot.png}
%\includegraphics[width = 60mm]{MF_Kpi.png}
%\vspace{-3mm}
%\caption{The $K^+/\pi^+$ ratios for PNJL, EPNJL models and PNJL model with vector interaction with $g_V=0.6 g_S$ for MFA and constant $\mu_s$.}
%\end{center}
%\label{nKnpi_ratio}
%\vspace{-5mm}
%\end{figure}

For the Beth-Uhlenbeck approach, lowering the pseudocritical temperature of the chiral restoration in EPNJL model  leads to a more rapid drop  above the horn, than in the BW approximation, when the chemical potential of pion is constant. 
%We introduced the nonequilibrium pion chemical potential and strange quark chemical potential. Now they are assumed to depend on the specific entropy variable $x = T/\mu_B$.
%\sout{ as a functions of the Woods-Saxon form.}
The pion chemical potential stems from the nonequilibrium character of pion production process which depends on the temperature and density of the system, so in particular in collisions with high pion multiplicities we can observe a frozen nonequilibrium state of the pions, with the nonequilibrium chemical potential $\mu_\pi$. It motivated us to introduce the specific functional form for its dependence on $x=T/\mu$. The strange quark chemical potential is a parameter which reflects the asymmetry in the production of positive and negative charged kaons. Since this imbalance is maximal at low energies and vanishes at high energies we adjusted a monotonously falling function of x, i.e. of $\sqrt(s)$. In order to find the x-dependence of $\mu_\pi$ and $\mu_s$  the following strategy was used: one can fit the pion chemical potential $\mu_pi$ at large $x$ which defines the overall normalization of ratio curves and error bar reflect the experimental uncertainty. Then we supposed that the pion chemical potential should be a non-decreasing function of $x$, while the ratio of strange chemical potential to nonstrange one should be decreasing one.  Therefore the $\mu_\pi(x)$ and $\mu_s(x)$ has a smooth behaviour in the transition region (which could be fitted with from the experimental points), and can be chosen as a functions of the Woods-Saxon form 

\begin{eqnarray}
\label{eq:mu-pi}
\mu_\pi(x) &=& \mu_\pi^{\mathrm{min}}  + \frac{\mu_\pi^{\mathrm{max}} - \mu_\pi^{\mathrm{min}}}{1 + \exp\left(-(x - x_\pi^{\mathrm{th}})/\Delta x_\pi\right) }, 
%\nonumber
\\
\mu_s(x) &=&\frac{ \mu_s^{\mathrm{max}}}{1 + \exp((x - x_s^{\mathrm{th}})/\Delta x_s) }.
\label{eq:mu-s}
\end{eqnarray}
The best values of the parameters for the PNJL (EPNJL) model are 
$\mu_\pi^{\mathrm{max}}= 147.6\pm10~(107\pm10)$ MeV, 
$\mu_\pi^{\mathrm{min}}=120~(92)$ MeV,
$x_\pi^{\mathrm{th}}=0.370~(0.409)$, 
$\Delta x_\pi = 0.015~(0.00685)$.
The parameter values
$\mu_s^{\mathrm{max}}/\mu_u^{\mathrm{crit}}= 0.205$, 
$x_s^{\mathrm{th}}=0.223$, $\Delta x_s = 0.06$
hold for both models.

\begin{figure}[!h]
	\centerline{
\includegraphics[width = 62mm]{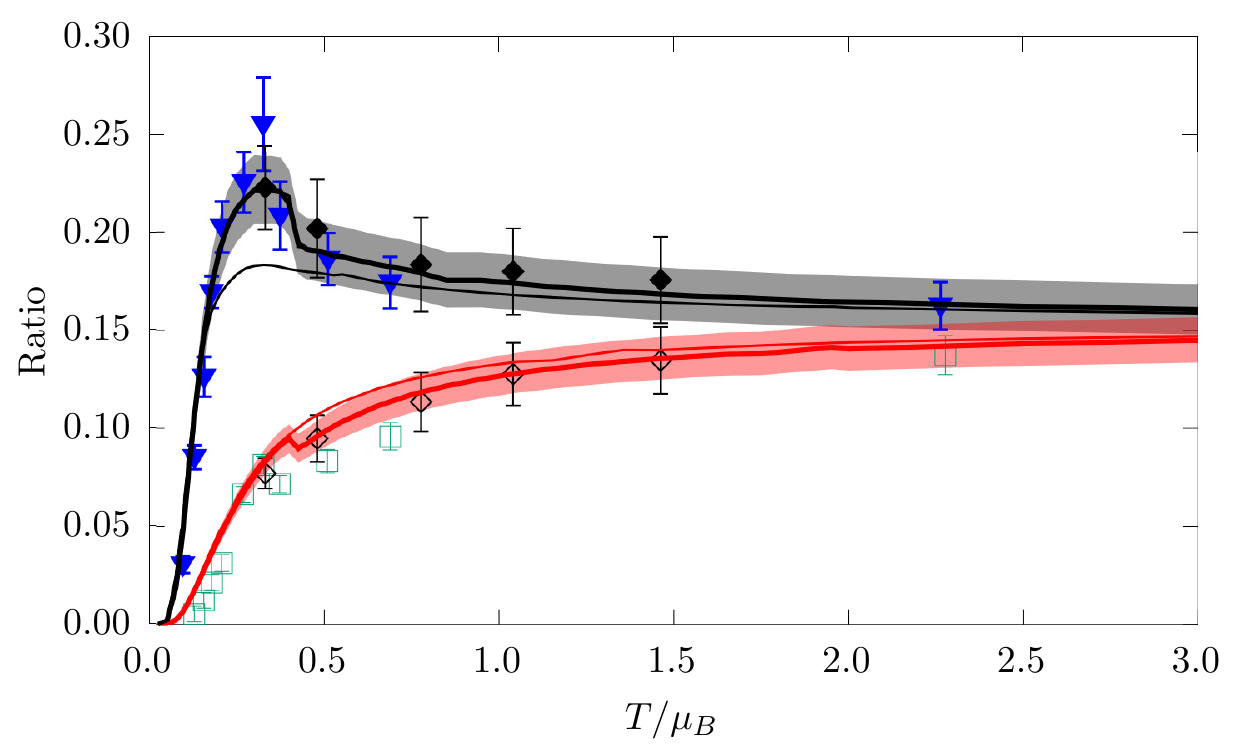}
\includegraphics[width = 62mm]{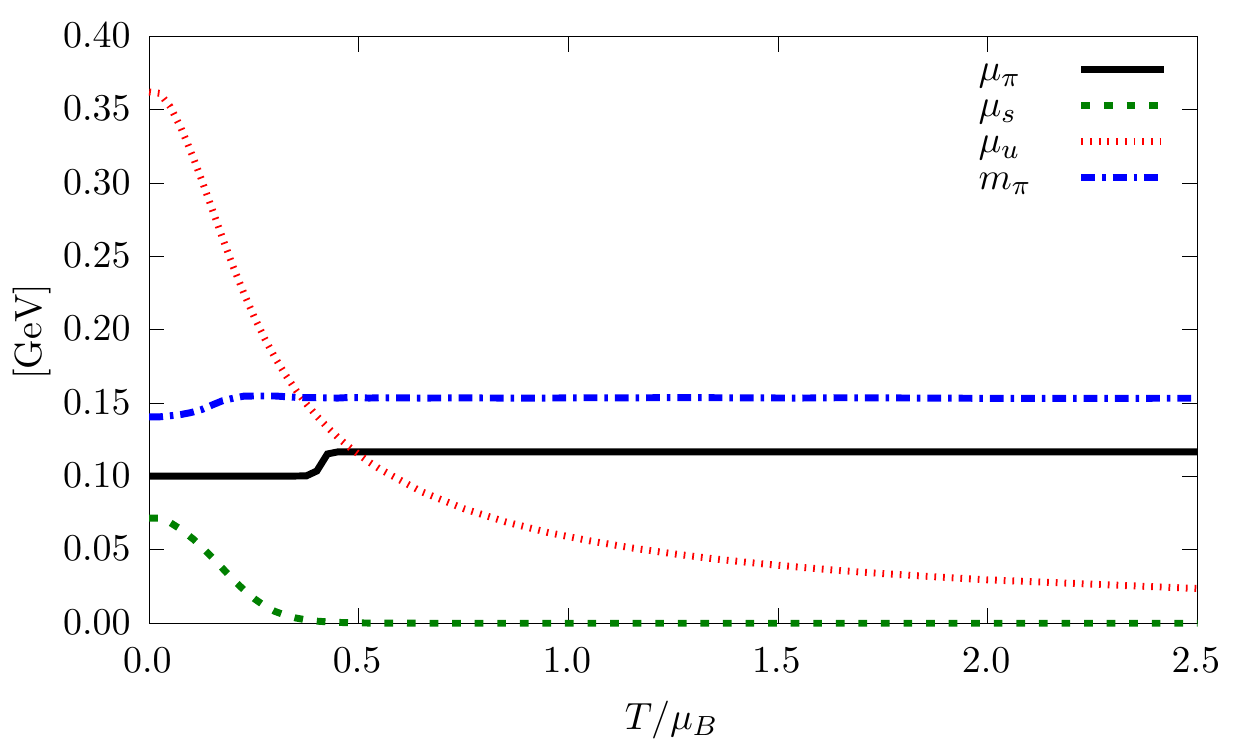}
	}
	\caption {Left panel: The ratios  $K^+/\pi^+$ (black lines) and $K^-/\pi^-$ (red lines) are shown as function of $T/\mu_B$ along the chemical freezeout line 	for the EPNJL model with $g_V=0$ within the BU approach. Thin lines correspond to the case with fixed $\mu_\pi=147.6$ MeV. The	thick lines are obtained when $\mu_s$ and $\mu_\pi$ vary with $T/\mu_B$ as shown in right panel. 
	}
	\label{nKnpi_ratioWithExpEnt}
\end{figure}

\begin{figure}[!h]
	\centerline{
\includegraphics[width = 63mm]{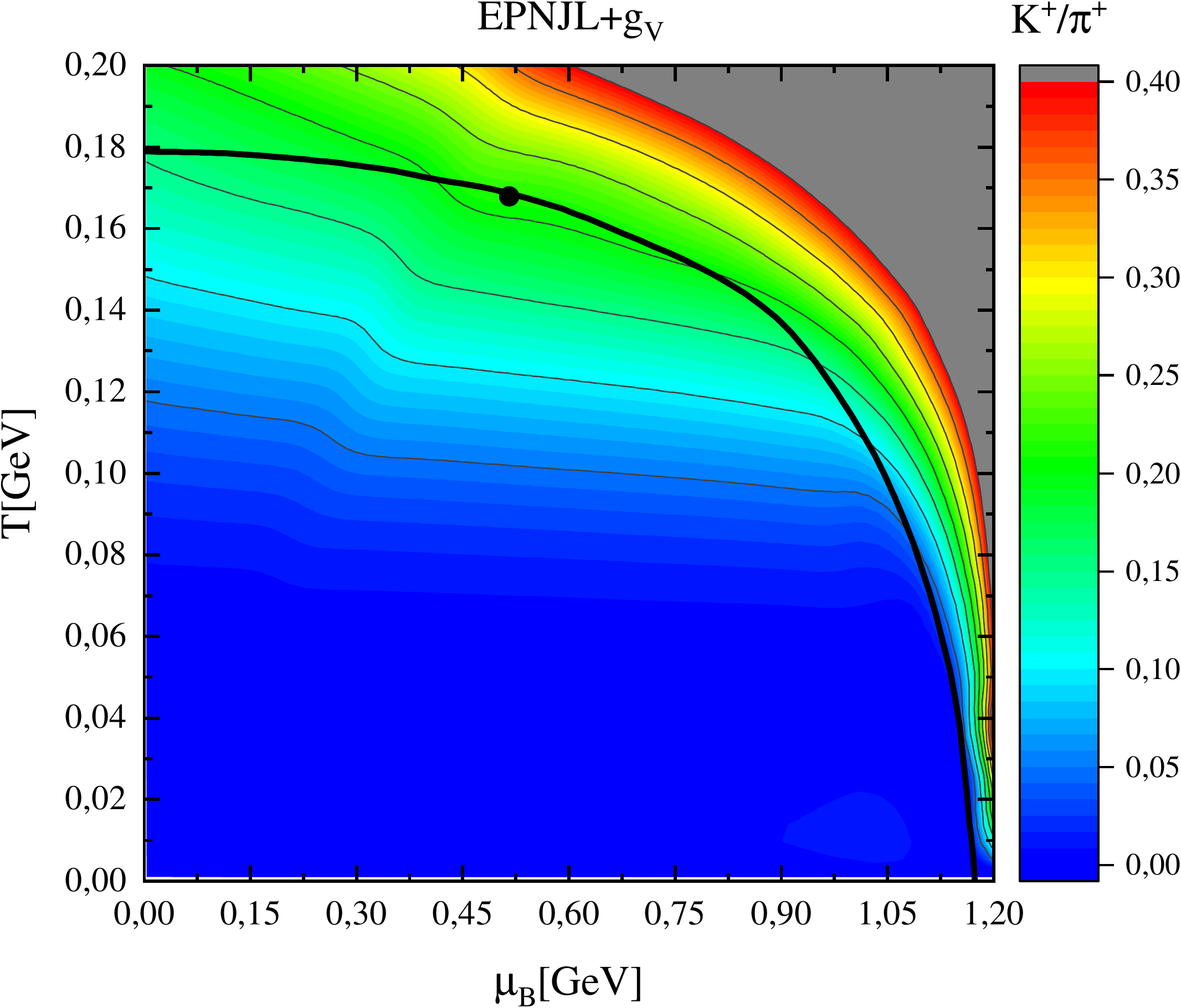}
\includegraphics[width = 58mm]{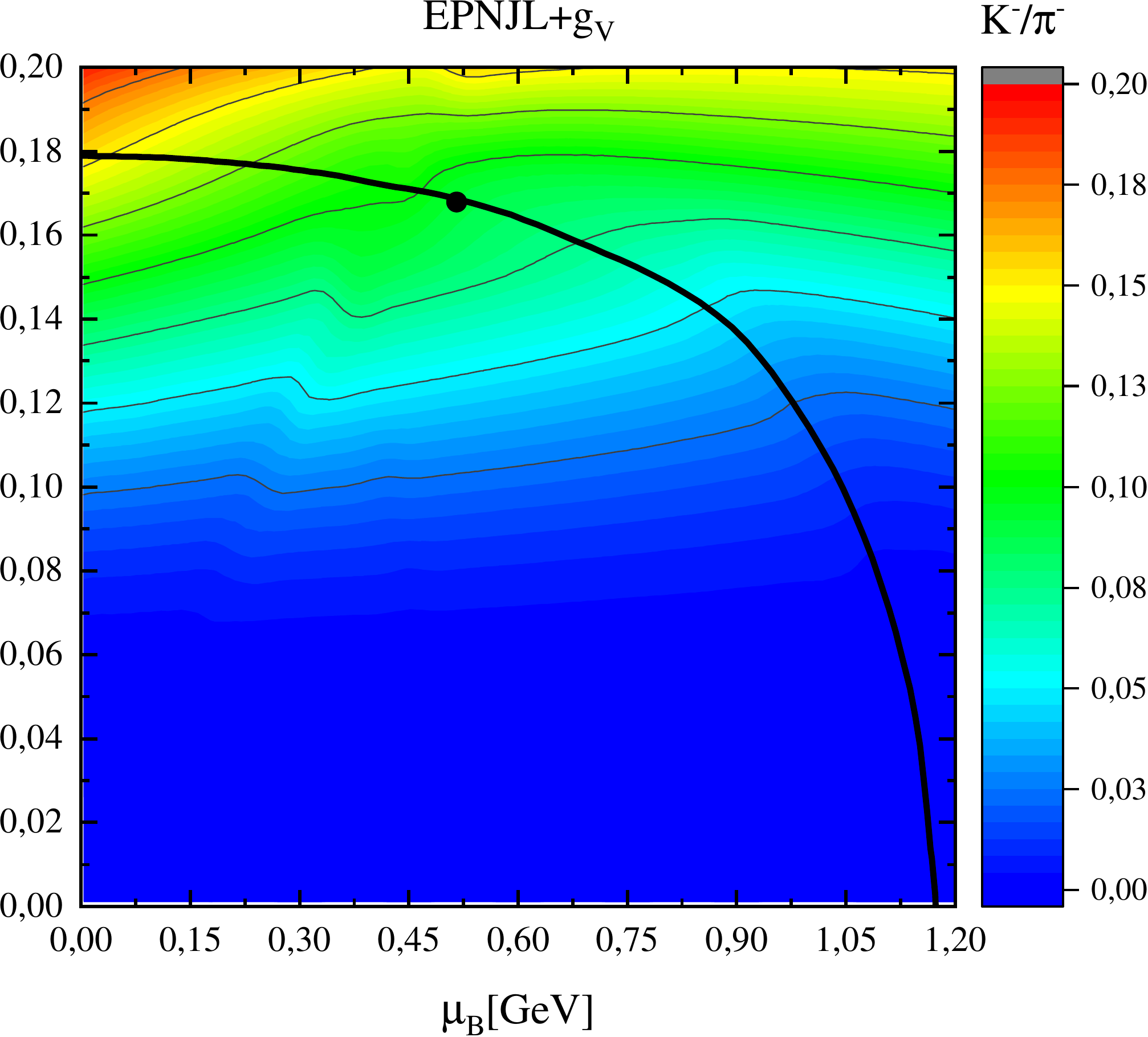}
	}
	\caption {$K^+/\pi^+$ (left) and $K^-/\pi^-$ (right) on the $T-\mu_B$ plane for the EPNJL model with $g_V = 0.6~g_S$ (no CEP). 
	The black dot indicates the maximum of $K^+/\pi^+$ ratio on the line of pseudocritical temperatures for the chiral transition (our proxy for chemical freeze out).
	}
	\label{Kpi_pd}
\end{figure}

In Fig. \ref{nKnpi_ratioWithExpEnt} we show the $K^+/\pi^+$  and $K^-/\pi^-$ for the EPNJL model as a function of the specific entropy variable $x=T/\mu_B$ when the BU approach is applied and the pion and strange quark chemical potentials are either constant (thin lines) or given as functions of $x$ (right panel) so that the experimental data can be fitted (thick lines).  To show the behaviour of $K/\pi$ ratios on the T-$\mu_B$, we choose the B.-Uh. approach with nonequilibrium pion chemical potential in EPNJL  model with $g_V=0.6 g_S$ (when there are no the first order chiral phase transition in the system). The dot on the phase diagram corresponds to the maximum of the $K^+/\pi^+$ ratio. It is clearly seen from the Fig. \ref{Kpi_pd} the smooth increasing of the value $K^-/\pi^-$ along the phase transition line (or with the increasing of energy, if we make a rescaling back). For the $K^+/\pi^+$ ratio the picture is more complicated: it shows the increase of the ratio in the middle part of the phase transition line.

\section*{Result and discussion}

In this work, we  used the SU(3) PNJL model to describe the QCD matter at finite temperature and density. In the frame of the model, we paid an attention to the $K/\pi$ ratio as a function of $T/\mu_B$, where T and $\mu_B$ are chosen along the phase transition line on the phase diagram. In our works we  replaced the $\sqrt{s_{NN}}$ by the variable $T/\mu_B$ choosing $T/\mu_B$ along the line of the phase transition  motivating that by the observation that along the straight lines of constant $x$ in the phase diagram the value of the specific entropy $S/N_B=s/n$ is constant and belongs to a conserved quantity in the hydrodynamic evolution of the hadronizing fireball created in central nucleus-nucleus collisions.

In the mean field PNJL model  we considered the cases with different values of $g_V$ which moves the CEP to lower T till it disappears and shifts the crossover pattern to higher chemical potentials.  According to our analysis, the key to understanding the horn lies in the position and the slope of the almost straight lines of constant $K^+/\pi^+$ and $K^-/\pi^-$ ratios relative to the curved freezeout line.  Both ratios are almost unaffected by the appearance of a  change from crossover to first order transition.

In the present work we have suggested that the sharpness of the horn effect in the $K^+/\pi^+$ ratio  is well explained by a Bose-enhanced pion production for $x>x_{\rm horn}$, i.e. for heavy-ion collisions with $\sqrt{s_{NN}}>8$ GeV, in the region of meson dominance, where also a low-momentum enhancement of pion production has been observed. 
Such an effect is best described by a nonequilibrium pion distribution function, which according to Zubarev's concept of the nonequilibrium statistical operator  requires an additional Lagrange multiplier, the pion chemical potential, for a consistent description \cite{Blaschke2020}. Within the Beth-Uhlenbeck approach we have provided fit functions for the $x$-dependence of the pion and strange quark chemical potentials that lead to a simultaneous description of the $x$-dependence of both kaon-to-pion ratios in accordance with the experimental data, with a strong horn effect for the $K^+/\pi^+$ ratio. 
In the present paper only the result of partial fit is presented. After finding the error bar of the  maximal value of the pion nonequilibrium chemical potential due to overall normalization to the large-$x$ points (with almost zero chemical potential) the other parameters is fitted. The detailed description of the fit and error bar estimation of all parameters procedure has been given in \cite{Blaschke:2020bzh}.

%\section*{Acknowledgements}
\vspace{0.2cm}

This work was supported by the Russian Fund for Basic Research (RFBR) under grant no. 18-02-40137. 
Numerical calculations are partially performed on computing cluster "Ac. V.M. Matrosov".

%\nocite{*}
%\bibliographystyle{pepan}
%\bibliography{maikbibl}
% Bibliography automatically generated via BibTeX (see template_bibtex.tex) 

\end{document}